\documentclass[11pt,twocolumn]{article}
\usepackage{graphicx}  
\usepackage{dcolumn}   
\usepackage{bm}     
\usepackage{adjustbox}
\usepackage{amsmath,amssymb}
\usepackage[caption=false]{subfig}
\usepackage{float}
\usepackage{makecell}
\usepackage[thinlines]{easytable}
\usepackage{array,booktabs}
\usepackage{lipsum} 
\usepackage{appendix}
\usepackage[colorlinks=true,linkcolor=blue,citecolor=blue]{hyperref}
\usepackage{ulem}
\usepackage{authblk}
\usepackage[total={7.0in, 10.0 in}]{geometry}
\usepackage[colorinlistoftodos]{todonotes} 
\usepackage[switch]{lineno} 

\newcolumntype{P}[1]{>{\centering\arraybackslash}p{#1}}

\title{Dynamical dark energy in the light of DESI 2024 data}
\author[1]{Nandan Roy}
\affil[1]{NAS, Centre for Theoretical Physics \& Natural Philosophy, Mahidol University, Nakhonsawan Campus, Phayuha Khiri, Nakhonsawan 60130, Thailand}
\affil[ ]{\texttt{nandan.roy@mahidol.ac.th (Corresponding Author)}}

\begin{document} 

\twocolumn[
  \begin{@twocolumnfalse}
    \maketitle
    \begin{abstract}

The latest findings from the DESI (Dark Energy Spectroscopic Instrument) data release 1 (DR1)~\cite{DESI:2024mwx}, combined with data from the cosmic microwave background and supernovae, suggest a preference for dynamical dark energy over the cosmological constant. This study has considered the Chevallier-Polarski-Linder (CPL) parameterization for the dark energy equation of state (EoS) and has indicated a possible phantom barrier crossing in the recent past. Despite CPL being the most commonly used parameterization, recent research has pointed out issues with its prior selection and parameter degeneracies. In this paper, we propose an alternative two-parameter parameterization of the dark energy equation of state (EoS). At higher redshifts, it behaves like the cosmological constant. At redshifts $z<1$, this parameterization closely approximates the CPL form but deviates from it at lower redshifts.  Our findings also indicate that the current value of the EoS of dark energy resembles quintessence, with evidence of a recent crossing of the phantom barrier, supporting the conclusions in ~\cite{DESI:2024mwx}. Furthermore, our model significantly reduces the Hubble tension to about $2.8\sigma$ when compared to Hubble Space Telescope and SH0ES data ~\cite{Riess:2021jrx}, and to $1.6\sigma$ with standardized TRGB and Type Ia supernova data ~\cite{scolnic2023cats}. Bayesian model selection using Bayes factors and Akaike Information Criteria (ACI),   shows a strong preference for our parameterization over the $\Lambda$ CDM model, aligned with the DESI2024 results and favoring dynamical dark energy.
\end{abstract}
  \end{@twocolumnfalse}
]

\maketitle



\section{Introduction}

A wide variety of cosmological observations ~\cite{SupernovaSearchTeam:1998fmf, SupernovaCosmologyProject:1998vns, Planck:2014loa, ahn2012ninth} have indicated that our universe is currently undergoing a mysterious accelerated expansion phase. No known form of matter can account for this accelerated expansion. Although the present observations align with the cosmological constant ~\cite{padmanabhan2006dark} as the driving force behind the accelerated expansion, significant theoretical issues have been encountered over time, such as the cosmological constant problem, the coincidence problem, etc., and an increasing number of observational challenges.

Recently, as our precision in the measurement of the cosmological parameters has improved significantly, the estimation of different cosmological parameters from different cosmological observations shows a discrepancy between their estimation. The most significant one is the Hubble tension. A statistically significant discrepancy of the order of $\simeq 5.3 \sigma$~\cite{Riess_2022} has been reported from the early universe measurements and the late time distance ladder measurement of the current value of the Hubble parameter ($H_0$). The early universe measurements like CMB from Planck collaboration ~\cite{Planck2020},  BAO ~\cite{BAO2017, BAO2011, DES2018,DES:2018rjw}, and BBN ~\cite{BBN2021} data estimated the current value of the Hubble constant to be $H_0 \sim (67.0 - 68.5)$ km/s/Mpc by considering the cosmological constant ($\Lambda$) as the candidate for dark energy. Whereas the measurement of $H_0$ using distance ladder measurements such as SH0ES ~\cite{Riess:2021jrx} and the collaborations of H0LiCOW ~\cite{Wong:2019kwg} reported that its value to be $H_0 = (73.04 \pm 1.42)$ km/s/Mpc and $H_0 = (73.3_{-1.8}^{+1.7}$ km/s/Mpc respectively. This discrepancy indicates that the dynamics of dark energy might be richer than simply a constant.  

Considering dark energy as dynamical is one approach to alleviating the challenges posed by the cosmological constant. A variety of dynamical dark energy models have been suggested, each presenting unique benefits and limitations~\cite{amendola2010dark, Bamba:2012cp, Mukhopadhyay:2024fch}. Despite more than two decades since the discovery of the accelerated expansion of the universe and extensive efforts to explain it, the underlying physics remains unresolved, with no clear consensus. Attempts have been made to reconstruct various cosmological parameters by parameterizing them and validating them against observational data. The parameterization technique is crucial as it allows us to model the unknown physics of dark energy effectively without altering the underlying theory of gravity. In this approach, a functional form is assumed for certain cosmological parameters, such as the equation of state (EoS) of dark energy, the Hubble parameter, or the dark energy density, and the model is constrained using state-of-the-art cosmological data.

The simplest dynamical dark energy model is the fluid description of the dark energy in which the dark energy equation of state is parameterized phenomenologically and the most popular parameterization of the dark energy EOS is the CPL parameterization or $w_0 w_a$ CDM model in which the EOS is parameterized as $w(a)= w_0 + w_a (1- \frac{a}{a_0})$. Recent DESI 2024 Data Release 1 (DR1) ~\cite{DESI:2024mwx} has constrained the CPL model and found the preference of more than $2\sigma$ for the dynamical dark energy over $\Lambda CDM$. Furthermore, findings from DESI indicate that dark energy recently transitioned from a phantom state to quintessence by crossing a phantom barrier\cite{Linder:2024rdj}. Additional research has presented comparable findings regarding the CPL model ~\cite{wang2024self,park2024using}. Conversely, ~\cite{Carloni:2024zpl} suggested that among numerous dynamic dark energy models analyzed, CPL is not the most suitable model for fitting the data. In ~\cite{cortes2024interpreting,shlivko2024assessing}, the authors have shown concern about the choice of prior and degeneracy, which can affect the result reported in ~\cite{DESI:2024mwx}. DESI 2024 data have also been used to constrain other cosmological models~\cite{DESI:2024kob,Bousis:2024rnb,Calderon:2024uwn,Yang:2024kdo,Wang:2024dka,Giare:2024smz,Berghaus:2024kra,Colgain:2024xqj,Dinda:2024kjf,Dinda:2024ktd,Reboucas:2024smm,Mukherjee:2024ryz} showing a preference for dynamical dark energy.

In this study, we propose an alternative parameterization of the dark energy equation of state (EoS) inspired by the dynamics of the quintom scalar field model~\cite{Roy:2023vxk}, which includes contributions from both quintessence and phantom scalar field, allowing us to further confirm the nature of dark energy as indicated by the DESI results~\cite{DESI:2024mwx}. An intriguing aspect of this proposed parameterization is its ability to resemble the CPL parameterization at redshifts $z<1$, while deviating from it at even lower redshifts. Additionally, at higher redshifts, it behaves like the cosmological constant. This suggests that our model may capture more complex behavior than the CPL parameterization in the low-redshift regime, while preserving the early-time physics by remaining consistent with the cosmological constant. We utilized the Pantheon Plus data with and without Chepids host distance anchor, along with the DESI BAO 2024 data and a compressed Planck likelihood, to constrain our model. Our results were compared with the $\Lambda$CDM model, and we applied Bayesian model selection using the Bayes factor to assess the preference for our model. In addition, we reported on the status of the Hubble tension in this context. To evaluate the performance of the proposed parameterization at the linear perturbation level, we computed the temperature anisotropies and the matter power spectrum (MPS), comparing them with those of the $\Lambda$CDM model.

 The current work is presented as follows: In Section \ref{sec:mathematicalbg}, we provide an overview of the mathematical formulation and discuss the proposed model. Section \ref{observation} discusses the observational data used in this work, along with the constraints obtained on the cosmological parameters. In Section \ref{sec:tension}, we compare the models and report the status of the Hubble tension. Finally, we conclude with our results and findings in Section \ref{conclusion}.

\section{Mathematical Setup and the Model} \label{sec:mathematicalbg}

In a spatially flat homogeneous and isotropic universe filled with barotropic fluids, the Friedmann constrain equation can be written as

\begin{equation}
\begin{split}
\frac{H^2(z)}{H_0^2}&= 
 \Omega_{\mathrm{m}, 0}(1+z)^3+\Omega_{\mathrm{r}, 0}(1+z)^4 \\
 &+\left(1-\Omega_{\mathrm{m}, 0}-\Omega_{\mathrm{r}, 0}\right) f_{\mathrm{DE}}(z),
\end{split}
\end{equation}

 where $\Omega_{m,0}$ and $\Omega_{r,0}$ are the current matter and radiation density respectively and $f_{\mathrm{DE}}(z) \equiv \rho_{\mathrm{DE}}(z) / \rho_{\mathrm{DE}, 0}$ is the normalized dark energy density given by the following expression,
 
  \begin{equation}
f_{\mathrm{DE}}(z) \equiv \frac{\rho_{\mathrm{DE}}(z)}{\rho_{\mathrm{DE}, 0}}=\exp \left(3 \int_0^z[1+w(\tilde{z})] \mathrm{d} \ln (1+\tilde{z})\right) .
\end{equation}

 Although considering any candidate for dark energy the expansion of the universe is affected by the evolution of this normalized dark energy density $f_{\mathrm{DE}}(z)$. Due to the unknown nature of dark energy, the most popular and simple way to investigate the dynamics of dark energy is to consider a parameterization of the dark energy equation of state. In this approach, there is no need to alter the underlying theory of gravity, and all the important cosmological parameters can be successfully reconstructed. Here we propose a new two-parameter parameterization of the dark energy equation of state in the following form,

\begin{equation} \label{eq1}
\begin{split}
 w_q(a) =&  {w_a} \left(\frac{a}{a_0}\right) Cos\left(\frac{a}{a_0}\right) \\
 &- {w_b} \left(\frac{a}{a_0}\right) Cosh\left(\frac{a}{a_0}\right) - 1
\end{split}
\end{equation}                            
The reason behind introducing the trigonometric and hyperbolic function in the equation of state is inspired by the equation of state of the scalar field dark energy models of quintessence and phantom fields. In ~\cite{Roy:2018nce,Urena-Lopez:2020npg,Roy:2018eug} it has been shown the equation of state of a quintessence field ($w_{DE} > -1$) can be written as a \textit{cosine} function whereas in ~\cite{LinaresCedeno:2021aqk} it has been shown the equation of state of a phantom scalar field ($w_{DE} < -1$) can be written as the function of the \textit{cosh} and for the quintom model it to be a function of the both \textit{cosine} and \textit{cosh} ~\cite{Roy:2023vxk}. The quintom models are known for its ability to cross the phantom divider line (PDL). Recent studies using model-independent approaches and data-driven approaches ~\cite{gerardi2019reconstruction, johri2004phantom,perkovic2019transient,Roy:2022fif,AlbertoVazquez:2012ofj,Escamilla:2021uoj,Gangopadhyay:2023nli,Sharma:2020unh} have shown the indication of PDL crossing for the dark energy sector. Also results from DESI (2024) ~\cite{DESI:2024mwx} have shown a preference for dynamical dark energy with a phantom barrier crossing. These results hint towards more complex late-time dynamics of the dark energy. 

The proposed parametrization consists of two dedicated parts: one corresponds to quintessence and the other to the phantom nature of the EoS. In cases $w_b=0$ and $w_a>1$  this EoS can show only quintessence-like behaviour since $0\leq Cos\left(\frac{a}{a_0}\right) \leq 1$ since $0\leq \frac{a}{a_0} \leq 1$. On the other hand, it can show only phantom-like behaviour when $w_a =0$ and $w_b>1$  since $ Cosh\left(\frac{a}{a_0}\right) \geq 1$. The existence of both these terms in EoS is expected to give rise to quintom-like behaviour with the possibility of smooth phantom barrier crossing. Another interesting feature of the proposed equation of state (EoS) is that for redshifts where \( \frac{a}{a_0} < 1 \), the approximations \( \cos\left(\frac{a}{a_0}\right) \approx 1 \) and \( \cosh\left(\frac{a}{a_0}\right) \approx 1 \) hold, in that regime, the parameterization can be well approximated by the Chevallier–Polarski–Linder (CPL) form:

\begin{equation}
    w_q(a) \simeq \tilde{w}_0 + \tilde{w}_a \left(1 - \frac{a}{a_0} \right),
\end{equation}

which is obtained by expanding the \( \cos \) and \( \cosh \) terms in a Taylor series. Here, the parameters are given by \( \tilde{w}_0 \simeq w_a - w_b -1 \) and \( \tilde{w}_a \simeq w_b - w_a \). 

At high redshifts, where \( \frac{a}{a_0} \ll 1 \), the equation of state asymptotically approaches \( w_q(a) \to -1 \), effectively mimicking the behavior of the cosmological constant. This behaviour of the $w_q(a)$ has been shown in Fig. \ref{fig:eos}.

We have implemented this proposed parameterization to the publicly available CLASS Boltzmann code~\cite{Lesgourgues:2011rg,Lesgourgues:2011rh} to numerically understand its behavior, and to constrain the model parameters against different cosmological observations, we have used the MontePython MCMC cosmological code~\cite{Brinckmann:2018cvx}.

\section{Observational Data}\label{observation}

To constrain the cosmological parameters, the following data sets have been used:

\subsection{Pantheon Plus with SH0ES R22}
Type Ia supernovae are widely used as standard candles because of their relatively uniform absolute luminosity~\cite{reiss1998supernova,SupernovaSearchTeam:1998fmf}. In this analysis, we used the Pantheon Plus compilation sample of SN-Ia data ~\cite{Brout:2022vxf,Riess:2021jrx} with and without the calibration from the Cepheid host distance anchors. This data set is often used to address the $M - H_0$ degeneracy in the supernova data. One can also use the Pantheon Plus data set with SH0ES prior on either $H_0$ or $M$. In \cite{Efstathiou:2021ocp,Camarena:2021jlr} it has been shown that one should use a prior on the $M$ rather than on $H_0$, since Cepheid calibrations are efficient for measuring the absolute magnitude. In Pantheon+SH0ES compilation, 
there are 1701 SNIa light curves data within which 1550 are in the redshift range $0.001 < z < 2.26$. Additionally, it combines 3 separate mid-z samples $(0.1 < z < 1.0)$, together with 11 different low-$z$ samples $(z < 0.1)$ and 4 separate high-z samples $(z > 1.0)$., Each of the separate data sets has its photometric systems and selection functions and provides values for the distance moduli $\mu$ at different redshifts.

\subsection{DESI BAO}
The density of visible baryonic matter exhibits recurring, periodic fluctuations called baryon acoustic oscillations. These oscillations are essential standard rulers for precise distance measurements in cosmology. In this study, we have used the 2024 BAO observation data from the Dark Energy Spectroscopic Instrument (DESI) as noted in reference ~\cite{DESI:2024mwx}. Covering a redshift range of $z\in [0.1,4.2]$, the survey is divided into seven redshift bins. BAO provides measurements of the effective distance along the line of sight as

\begin{equation}
\frac{D_H(z)}{r_d} =\frac{c r_d^{-1}}{H(z)}
\end{equation}

and along the transverse line of sight as,
\begin{equation}
    \frac{D_{M}(z)}{r_{d}}\equiv\frac{c}{r_{d}}\int_{0}^{z}\frac{d\tilde{z}}{H(\tilde{z})}=\frac{c}{H_{0}r_{d}}\int_{0}^{z}\frac{d\tilde{z}}{h(\tilde{z})}.
\end{equation}
The angle average distance is measured as,

\begin{equation}
\frac{D_V(z)}{r_d} =\left[\frac{c z r_d^{-3} d_L^2(z)}{H(z)(1+z)^2}\right]^{\frac{1}{3}}.
\end{equation}

Here, the luminosity distance is represented by $d_L(z)$. The effective redshift and the corresponding $D_M / r_d$, $D_H / r_d$, and $D_V / r_d$ ratios used in this work are from TABLE: 1 of reference ~\cite{DESI:2024mwx}. This data contains five different samples, the Bright Galaxy Sample(BGS), Luminous Red Galaxy Samples (LRG), Emission Line Galaxy Sample (ELG), Quasar Sample (QSO) and Lyman-$\alpha$ Forest Sample (Ly$\alpha$).


\subsection{Compressed Planck likelihood}
In this analysis, we have used the Planck compressed likelihood following the approach in ~\cite{Arendse_2020} (hereafter P18). While having limited computational resources this compressed likelihood is useful to obtain the same constraining accuracy as full Planck.  The compressed likelihood uses the baryon physical density $\omega_b = \Omega_b h^2$ and the two shift parameters $\theta_*=r_s\left(z_{\text {dec }}\right) / D_A\left(z_{\text {dec }}\right), \quad \mathcal{R}=\sqrt{\Omega_M H_0^2} D_A\left(z_{\text {dec }}\right),
$ where, the redshift decoupling is $z_{d e c}$, and $D_A$ is the comoving angular diameter distance. Here we chose the mean values of the above parameters to be $100 \omega_b=2.237\pm0.015,100 \theta_s=1.0411\pm0.00031$, and $\mathcal{R}=1.74998 \pm 0.004$, and the correlation matrix given in Appendix A of ~\cite{Arendse_2020}. 

Given the data sets above, following combinations of the data have been considered in this analysis:

\begin{itemize}
    \item Set 1: P18 $+$ DESI $+$ Pantheon Plus
    \item Set 2: P18 $+$ DESI $+$ Pantheon Plus $+$ SH0ES
\end{itemize}

\subsection{Observational Constraints}\label{sec:bestfit}

\begin{table*}[h!]
\centering
\begin{adjustbox}{width=\textwidth,center}
\begin{tabular}{|l|c|c|c|c|c|c| }
\hline
Data&\multicolumn{3}{c|}{Set 1: P18 + DESI + Pantheon Plus}& \multicolumn{3}{c}{Set 2: P18 + DESI + Pantheon Plus + SH0ES} \vline\\
\hline
\hline
Parameters & $\Lambda$ CDM & $w_q$CDM &CPL& $\Lambda$ CDM & $w$CDM & CPL\\
\hline
{\boldmath$\omega_{b}$} &  $0.02244\pm 0.00014            $ & $0.02241\pm 0.00015            $ & $0.02239\pm 0.00016            $ & $0.02267^{+0.00013}_{-0.00014}   $ & $0.02244^{+0.00017}_{-0.00012}   $  & $0.02243\pm 0.00015            $\\
{\boldmath$\omega_{cdm}$} & $0.11877\pm 0.00086 $& $0.1191\pm 0.0011          $ & $0.1194\pm 0.0013          $ & $0.11679^{+0.00095}_{-0.00071}$ & $0.11931^{+0.00077}_{-0.0013}$ & $0.1194\pm 0.0012          $\\
{$H_0$} & $67.69\pm 0.41$ &$67.44\pm 0.74             $ & $67.59\pm 0.73             $ & $68.73^{+0.32}_{-0.47}     $ & $69.56\pm 0.63             $ & $69.77\pm 0.61             $ \\
{\boldmath$\Omega_{DE}$} &$0.6903\pm 0.0053          $& $0.6873\pm 0.0073          $ & $0.6880^{+0.0074}_{-0.0066}$& $0.7032^{+0.0044}_{-0.0055}$ & $0.7056^{+0.0061}_{-0.0052}$ & $0.7071\pm 0.0057          $\\
{\boldmath$\Omega_m$}& $0.3096\pm 0.0053          $ & $0.3127\pm 0.0073          $ & $0.3119^{+0.0066}_{-0.0074}$ & $0.2967^{+0.0055}_{-0.0044}$ & $0.2943^{+0.0052}_{-0.0061}$ & $0.2928\pm 0.0057          $ \\
{\boldmath$w_a/\tilde{w}_0 $}& - & $-0.70\pm 0.40             $ & $-0.850\pm 0.080           $ & - & $-1.27^{+0.45}_{-0.37}     $ & $-0.838\pm 0.073           $\\
{\boldmath$w_b/\tilde{w}_a $}& - & $-0.37\pm 0.19             $ & $-0.61\pm 0.33             $& - &$-0.59\pm 0.20             $& $-0.98^{+0.34}_{-0.29}     $ \\
{\boldmath$w_{de}$} & -1& $-0.808^{+0.099}_{-0.089}  $ &   $-0.850\pm 0.080           $ & -1 &$-0.784^{+0.10}_{-0.089}   $ & $-0.838\pm 0.073           $\\
\hline
\end{tabular}
\end{adjustbox}
\caption{The constraint  ($68\%$ CL) obtained on cosmological parameters from the combining data sets 1 and 2. \label{Tab:constraint}}
\end{table*}

 We have considered flat priors on the cosmological parameters, $100~\omega_{b}:[0,4.6], \omega_{cdm}:[0.095,0.145]$. For the model parameters, we have considered the prior to be $w_{a}:[-3,1]$ and $w_{b}:[-1,1]$. Please see Appendix \ref{app:wz} where we have discussed how this choice of prior can incorporate our different classes of evolution off the $w_q(z)$. The mean value and the corresponding $68\%$CL constraint obtained on the cosmological parameters of our analysis are given in Table\ref{Tab:constraint} for both the combinations of the data provided in Set 1 and Set 2, where $w_q$CDM model represents the proposed parameterisation here.
 

 For comparison, the constraint on the $\Lambda CDM$ model and CPL model has been given together with the proposed model as $w_q$CDM model.

\begin{figure*}[h!]
    \centering
    \includegraphics[width=\textwidth]{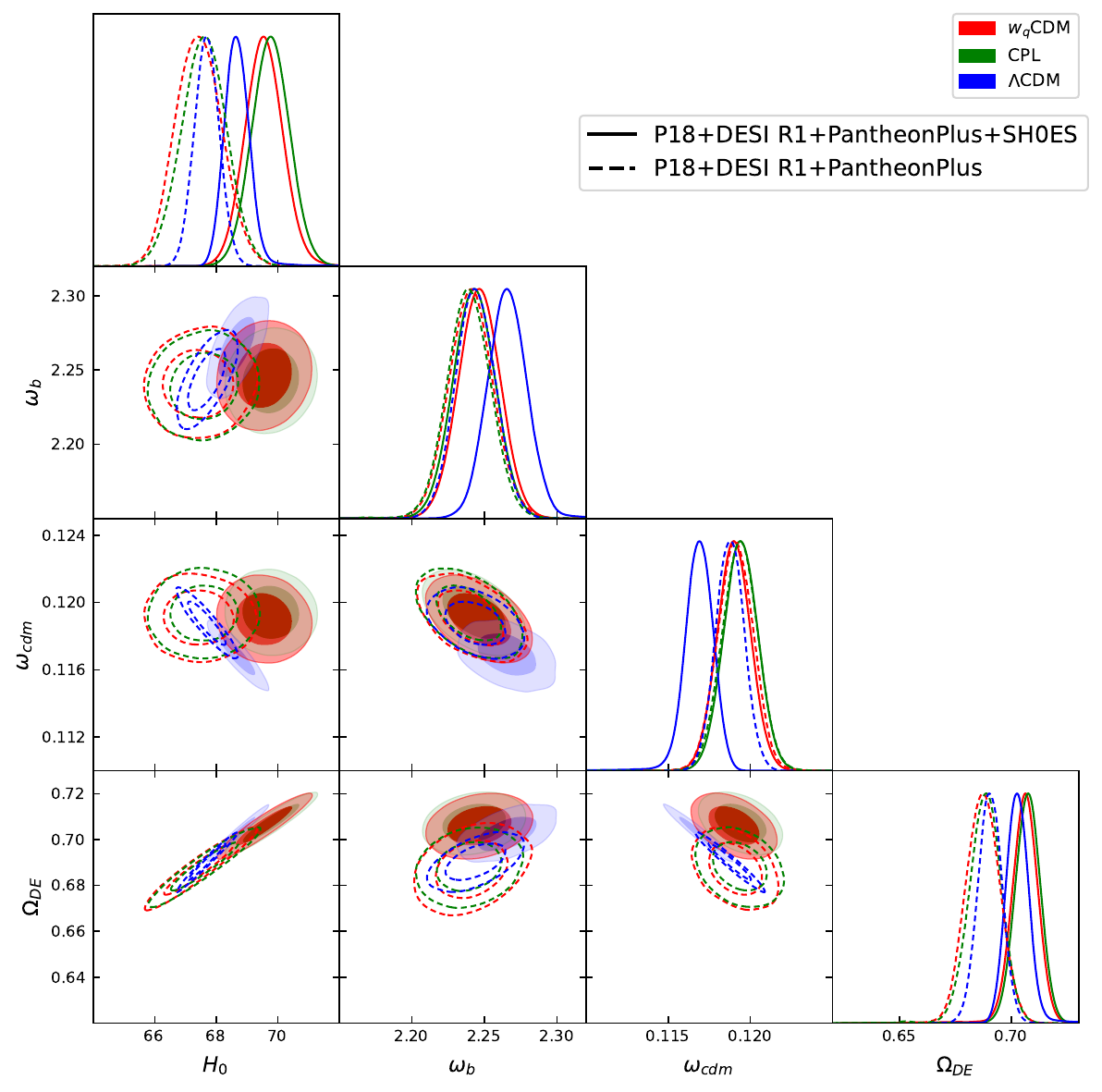}
    \caption{A triangular plot illustrates the constraints on cosmological parameters for the proposed model $w_q CDM$ (in red), with the $\Lambda$CDM model (in blue) and CPL model (in green) for comparison.  }
    \label{fig:cosmo}
\end{figure*}

  The 2D and 1D triangular plots of the cosmological parameters $w_b, w_{cdm},H_0, \Omega_{DE}$ are shown in FIG.\ref{fig:cosmo}. The posteriors for $\Lambda CDM$ are shown in blue, the CPL model in green and the $w_q$CDM model in red. The plots in the dashed show constraints from the combined data Set 1 without SH0ES data and the plots in the solid show the constraints from Set 2 with SH0ES data.

\begin{figure}[h!]
    \centering
\includegraphics[width=\columnwidth]{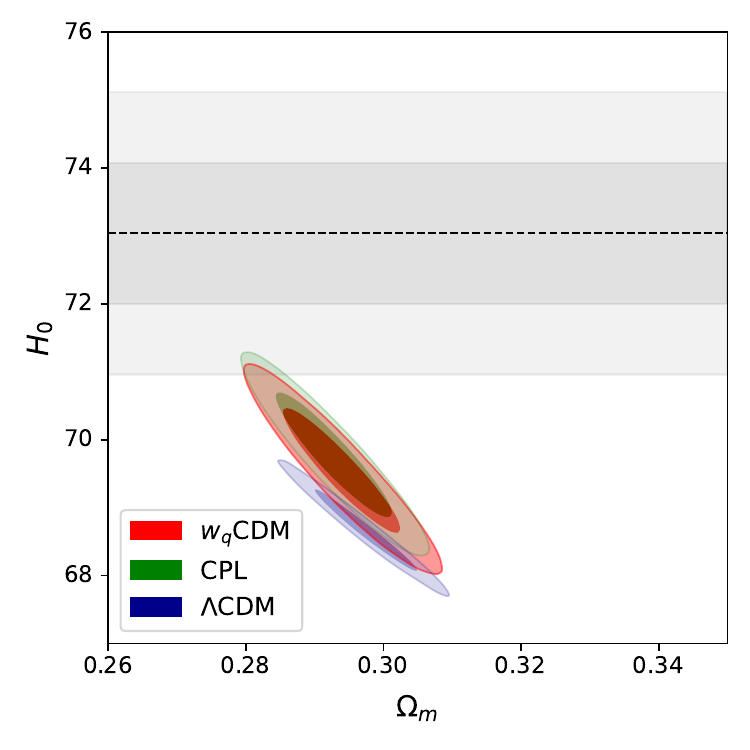}
    \caption{The figure shows two-dimensional posterior plots on $(\Omega_0,H_0)$ plane. The value of the $H_0$ from SH0ES collaboration~\cite{Riess:2021jrx} is shown by the green band. The deep and light grey regions represent $1\sigma$ and $3\sigma$ constraints respectively. }
    \label{fig:HT}
\end{figure}

  It can be seen that due to the inclusion of the SH0ES data, the value of the $H_0$ increases significantly for all the models. The proposed $w_q$CDM model suggests a higher value of $H_0$ and $\Omega_{DE}$ compared to $\Lambda CDM$ whereas the CPL model has the highest value of the $H_0$. In FIG.\ref{fig:HT} we have shown the 2D posterior distribution of $H_0$ versus $\Omega_m$ of $w_q$CDM model together with $\Lambda CDM$ and CPL model for comparison. The deep and light grey bands show, respectively, the constraints $1\sigma$ and $3\sigma$ on $H_0$ from ~\cite{Riess:2021jrx}.

\begin{figure*}
    \centering
\includegraphics[width=\textwidth]{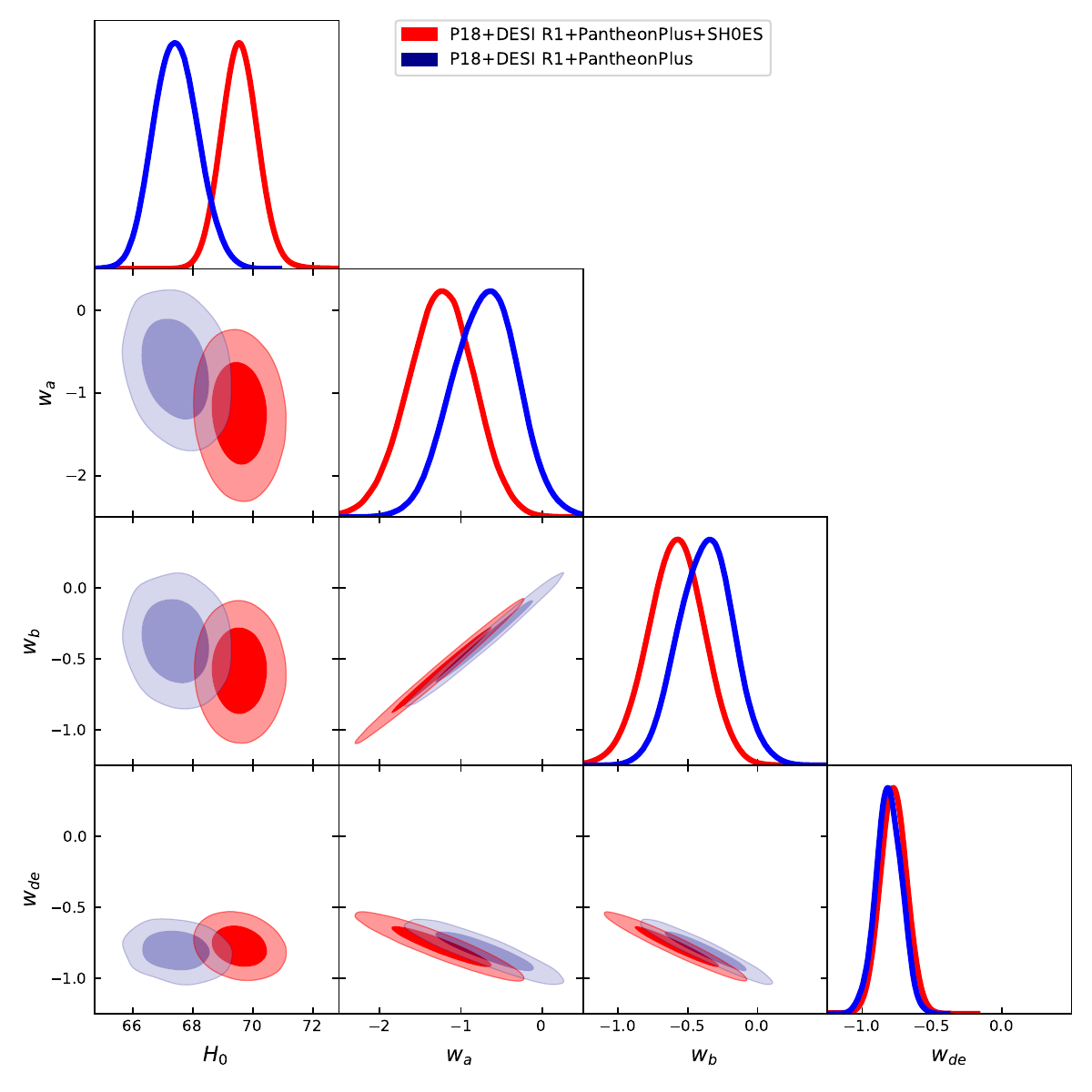}
    \caption{A triangular posterior plot showing the constraint obtained on the model parameters $w_a,w_b$ and the dark energy EOS ($w_{DE}$). }
    \label{fig:model}
\end{figure*}

\begin{figure}
    \centering
    \includegraphics[width=\columnwidth]{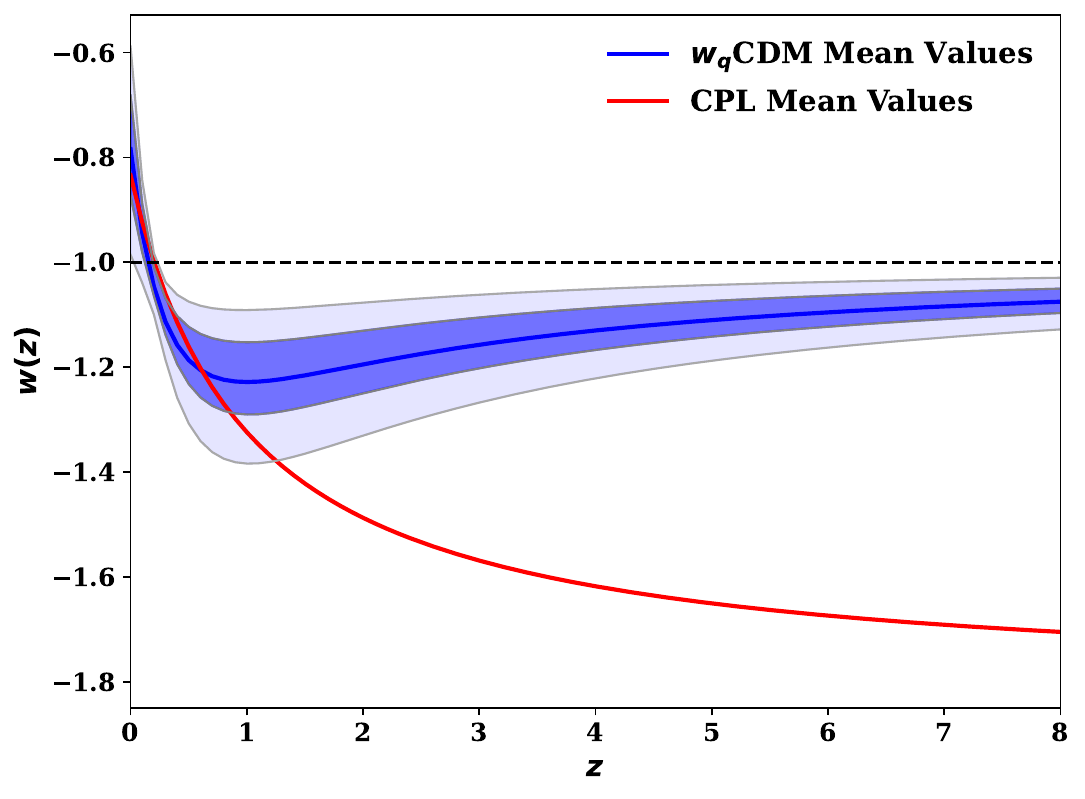}
    \caption{ This figure shows the posterior probability $Pr(w|z)$ of the dark energy equation of state (EOS). The dark blue area indicates the $1\sigma$ confidence level, while the light blue areas show the $2\sigma$ confidence level for the $w_q$CDM model. The evolution of the $w(z)$ vs $z$ for the CPL model has been shown in red.}
    \label{fig:eos}
\end{figure}

Figure~\ref{fig:model} shows the contour plot of posterior distributions of the model parameters $w_a, w_b$ and the current value of the dark energy EOS $w_{DE}$. The plots in blue and red are for Set 1 and Set 2 data combinations respectively. The current value of the EOS for all data combined (Set 2) is $w_{DE}=-0.784_{-0.089}^{+0.10} $ which is quintessence in nature. To understand the evolution of the EOS of the dark energy in FIG.\ref{fig:eos} we have plotted the posterior probability $Pr(w|z)$ of the $w_{DE}$ against the redshift $z$ with Set 2 data combination. The deeper and lighter blue region shows the contour level $1\sigma$ and $2\sigma$, while the solid blue line shows the mean value. However, it can be seen that the current value of the EOS is in the quintessence region, but in the recent past, it has a phantom barrier crossing around $z\approx 0.35 $ from phantom to quintessence. This finding of ours matches the one reported in ~\cite{DESI:2024mwx,desi2}. For comparison we have also plotted the evolution of the EOS of the dark energy for the CPL parametrization in red for the min value of the $\tilde{w}_0, \tilde{w}_a$ parameters given in Table \ref{Tab:constraint} for the Set 2 data combinations. Both the parametrization shows very similar evolution in the $z < 1$ range and differ as $z$ approaches smaller values. At the higher redshift, the EOS for the $w_q$CDM asymptotically approaches $-1$ and hence can mimic the cosmological constant, whereas, for the CPL model, it approaches deeper in the phantom region.

\begin{figure}[h!]
    \centering
    \includegraphics[width=1.0\columnwidth]{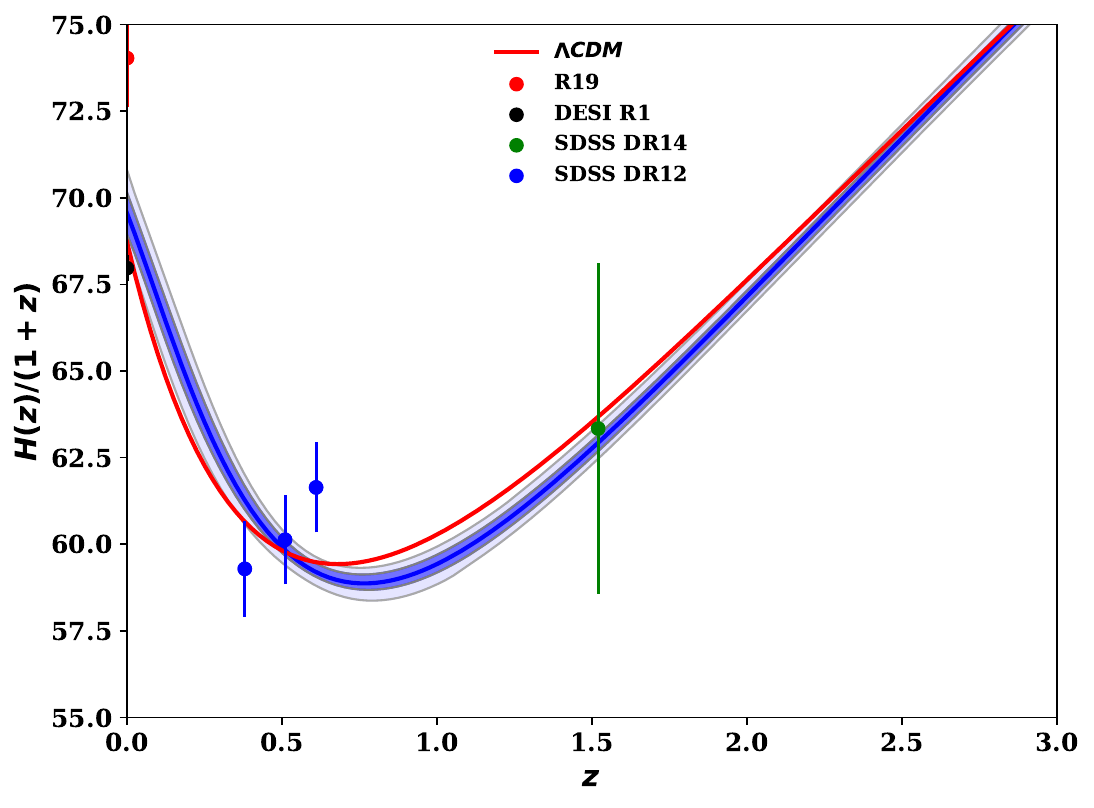}
    \caption{This figure shows $Pr(H(z)/(1+z))$ versus $z$ for the current model, compared with Sh0ES survey data ~\cite{Riess:2019cxk} and BAO surveys ~\cite{BOSS:2016wmc,Zarrouk:2018vwy,Blomqvist:2019rah,deSainteAgathe:2019voe}. The dark blue area represents the $1\sigma$ confidence level, and the light blue areas represent the $2\sigma$ confidence level. The red line shows the prediction from the $\Lambda$CDM model.}
    \label{fig:H(z)}
\end{figure}

\begin{figure}[h!]
    \centering
    \includegraphics[width=\linewidth]{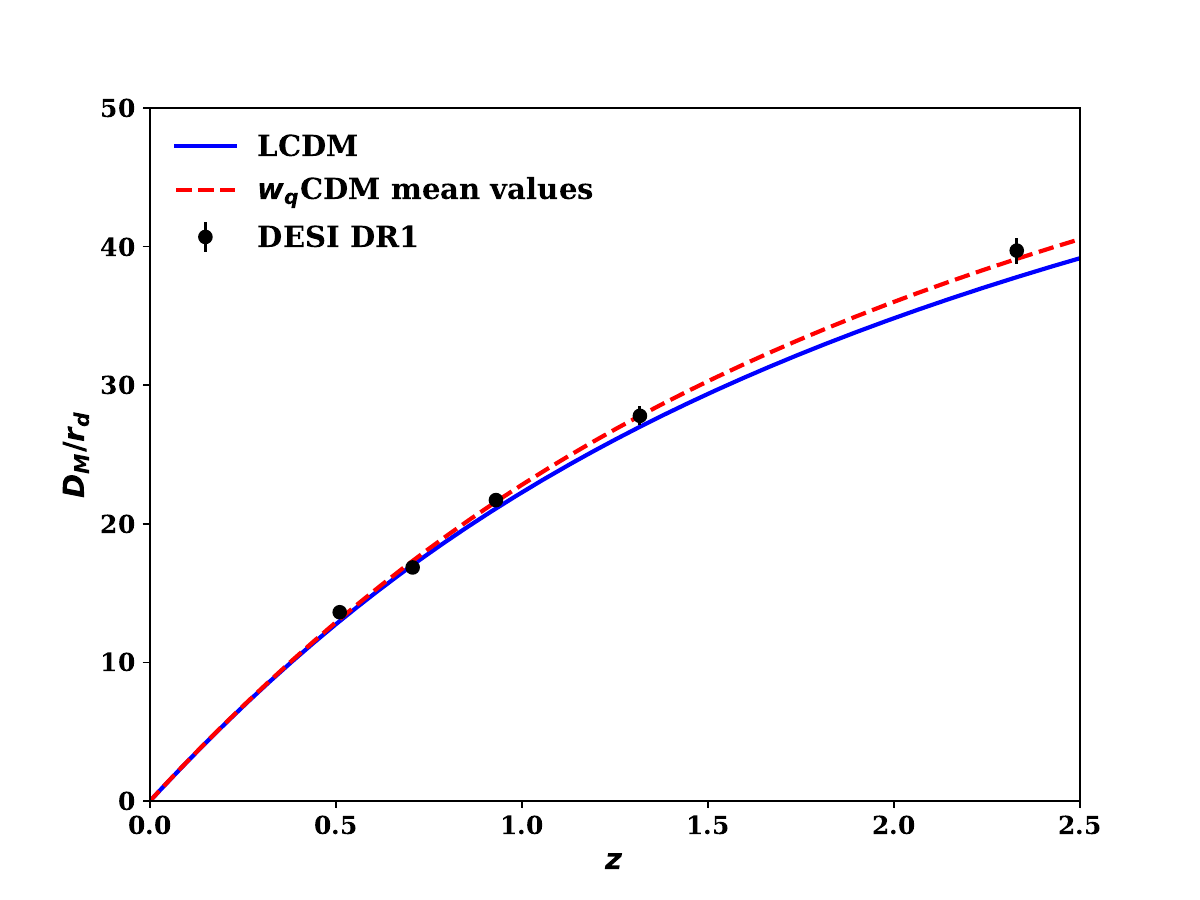}
    \caption{The evolution of the $D_M/r_d$ with respect to redshift $z$ for the $w_q$CDM model is represented in red. The DESI DR1 BAO data points are indicated by black dots, along with the predictions from the $\Lambda$CDM model for comparison.}
    \label{fig:dm}
\end{figure}

\begin{figure}[h!]
    \centering
    \includegraphics[width=\linewidth]{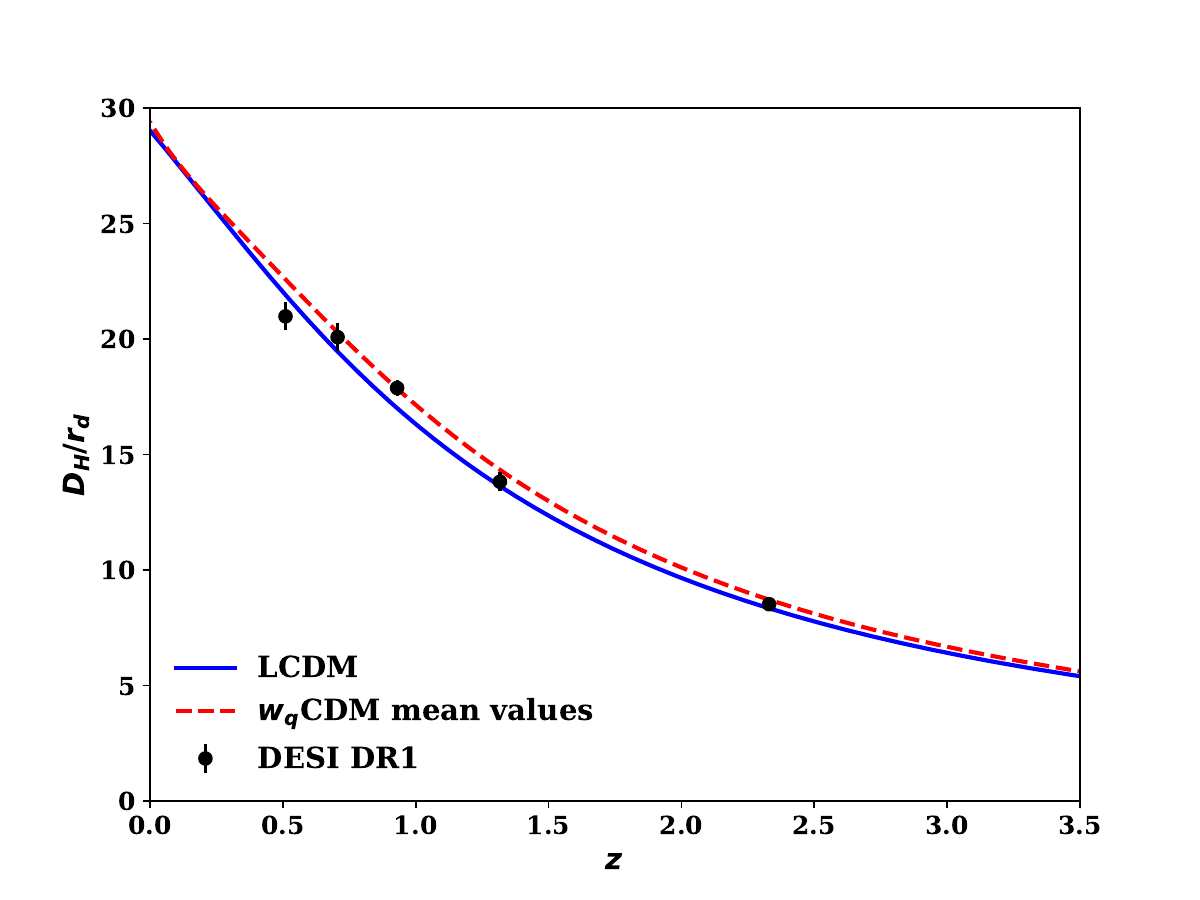}
    \caption{The evolution of the $D_H/r_d$ with respect to redshift $z$ for the $w_q$CDM model is represented in red. The DESI DR1 BAO data points are indicated by black dots, along with the predictions from the $\Lambda$CDM model for comparison.}
    \label{fig:dh}
\end{figure}

\begin{figure}[h!]
    \centering
    \includegraphics[width=\columnwidth]{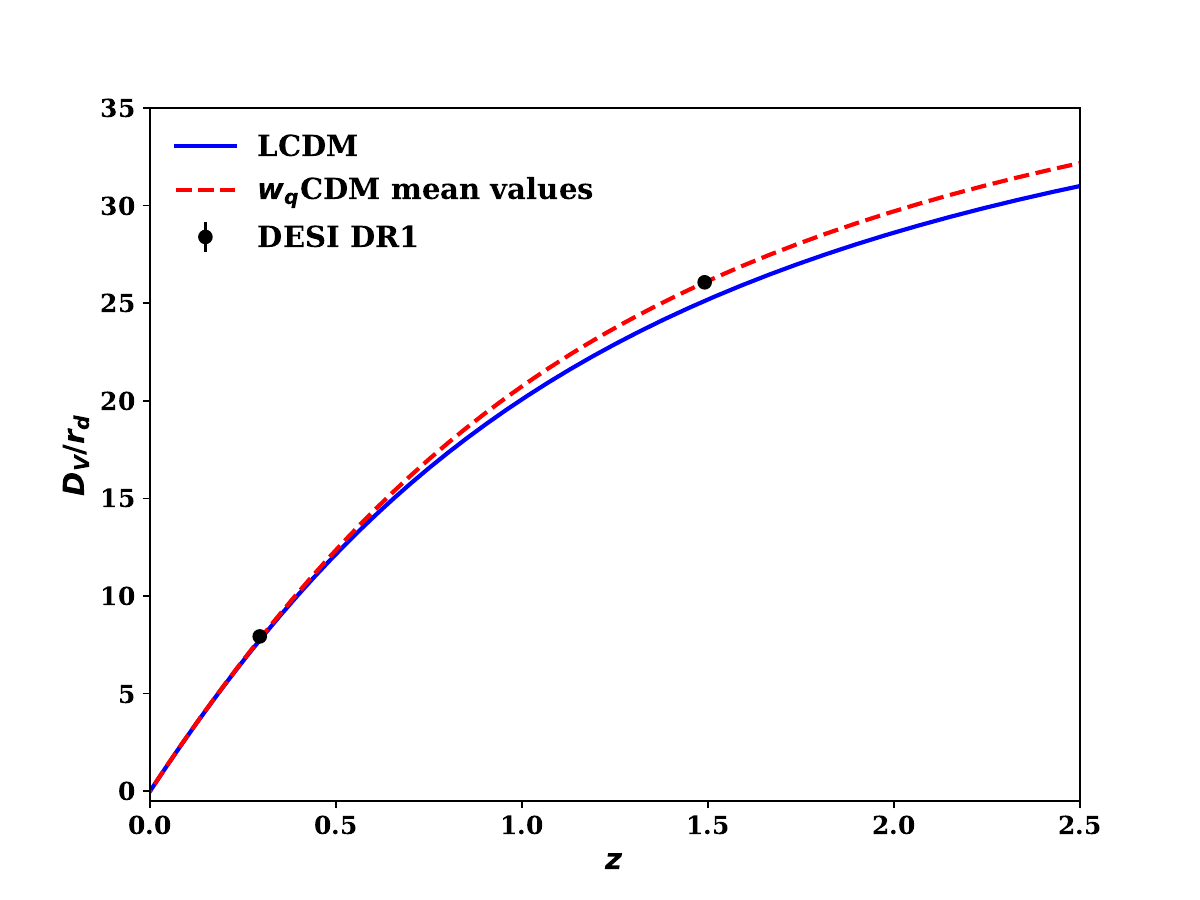}
    \caption{The evolution of the $D_V/r_d$ with respect to redshift $z$ for the $w_q$CDM model is represented in red. The DESI DR1 BAO data points are indicated by black dots, along with the predictions from the $\Lambda$CDM model for comparison.}
    \label{fig:dv}
\end{figure}

Although we have obtained constraints on the main cosmological parameters as expected, it is necessary to verify the evolution of the Hubble parameter against the observed data. Therefore, we have plotted the posterior probability of $H(z)/(1+z)$ vs. $z$ in FIG. \ref{fig:H(z)} for the mean value of the parameters obtained from data combination Set 2, along with observational data points from the SH0ES survey ~\cite{Riess:2019cxk} and Baryon Acoustic Oscillation (BAO) surveys ~\cite{BOSS:2016wmc,Zarrouk:2018vwy,Blomqvist:2019rah,deSainteAgathe:2019voe, DESI:2024mwx}. The evolution of $H(z)/(1+z)$ shows the expected behavior of the proposed model. The $H(z)/(1+z)$ for the $\Lambda$CDM model is shown by the red line for comparison. In Fig. \ref{fig:dm}, Fig. \ref{fig:dh} and Fig. \ref{fig:dv} we have plotted the evolution of the $D_M/r_d, D_H/r_d$ and $D_v/r_d$ for the $w_q$CDM model in red together with the observed data from DESI DR1 \cite{DESI:2024mwx}. The prediction from the $\Lambda$CDM is shown in blue for comparison.

\begin{figure}[h!]
    \centering
    \includegraphics[width=\columnwidth]{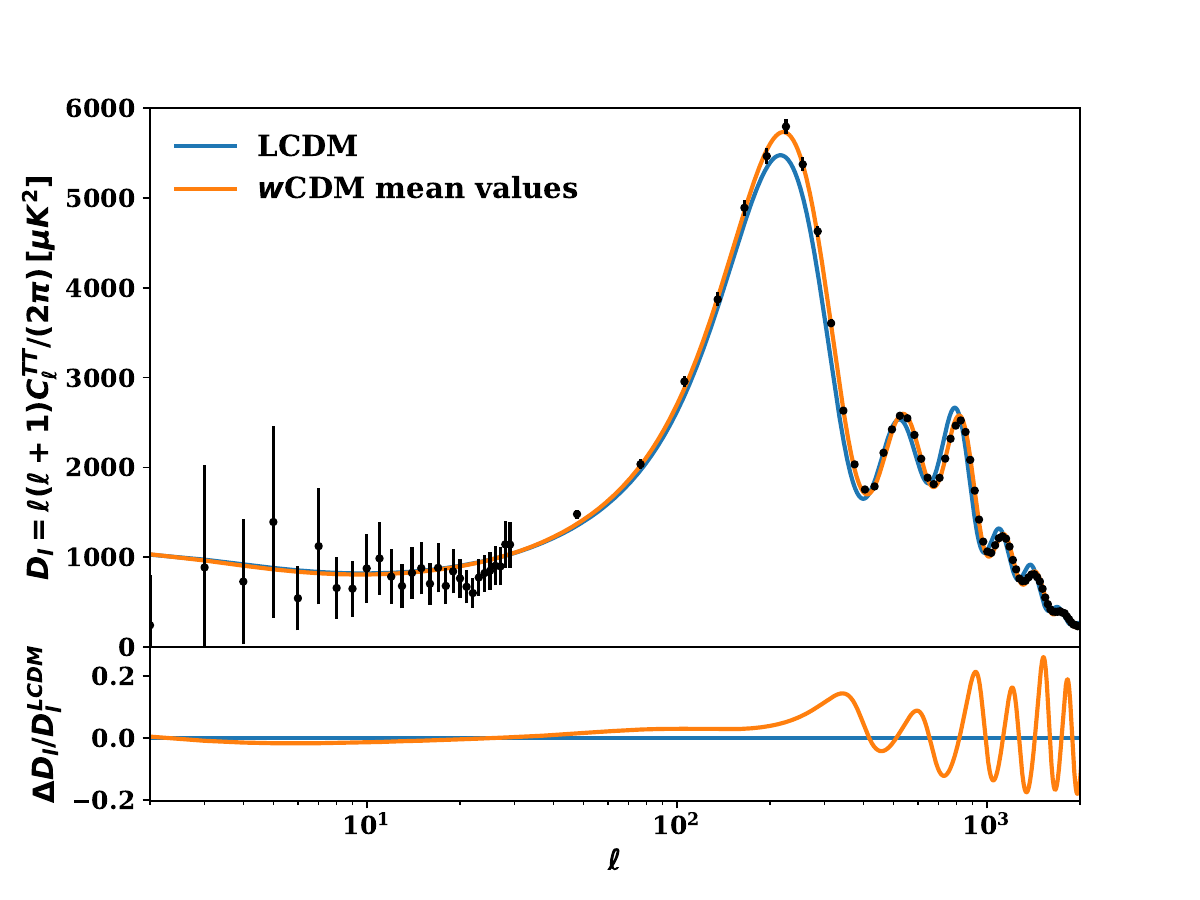}
    \caption{Plot of the CMB anisotropies for the $w CDM$ model (in yellow) using the mean value of the model parameters $w_a, w_b$ together with the $\Lambda CDM$ model (in blue). The bottom panel shows the relative difference between the models $\triangle D_l=\left(D_l-D_i^{\Lambda C D M}\right) / D_l^{\Lambda C D M}$.}
    \label{fig:dl}
\end{figure}

\begin{figure}[h!]
    \centering
    \includegraphics[width=\columnwidth]{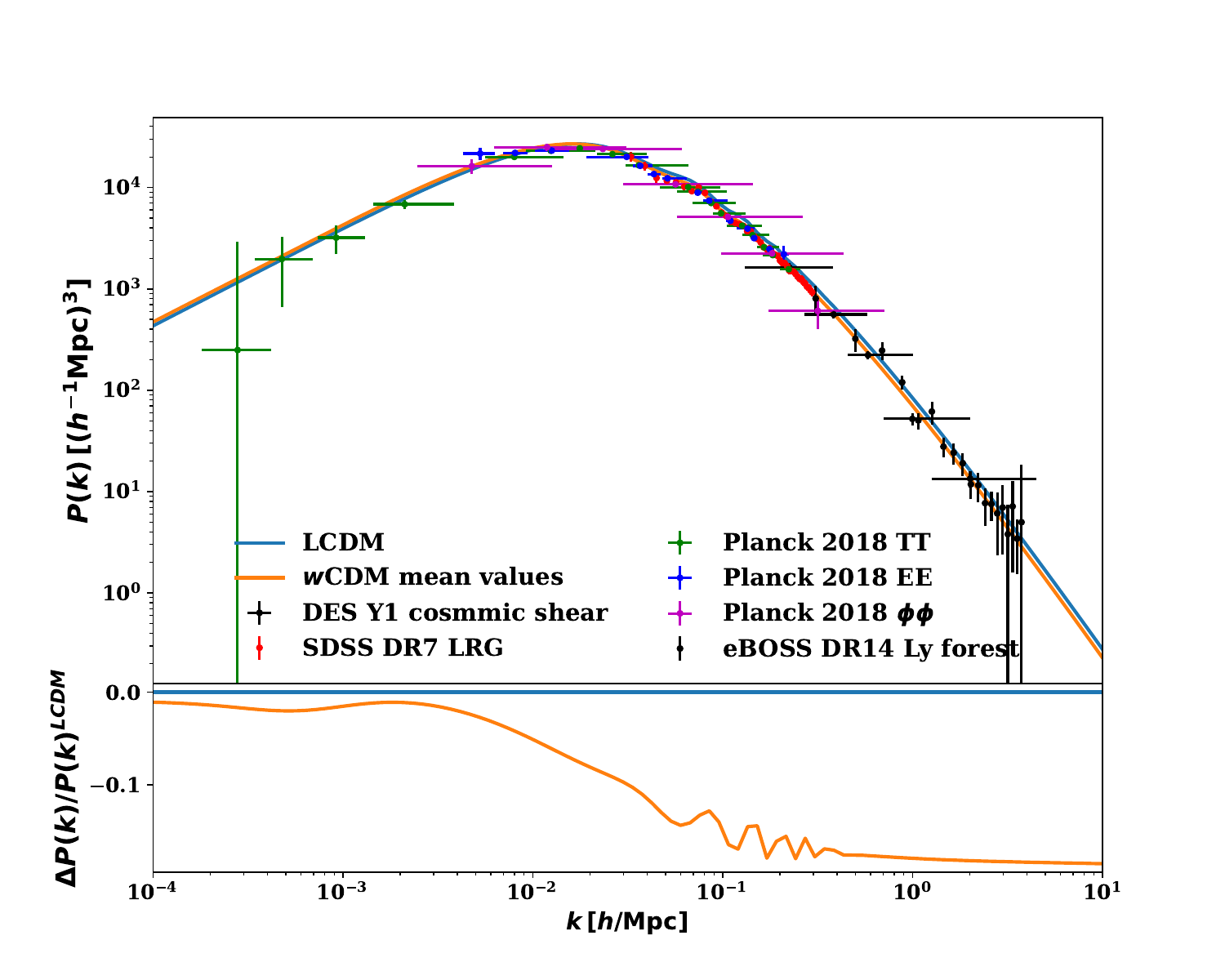}
    \caption{The Plot of the matter power spectrum for the $w CDM$ model (in yellow) using the mean value of the model parameters $w_a, w_b$ together with the $\Lambda CDM$ model (in blue). The bottom panel shows the relative difference between the models $\triangle P(k)=\left(P(k)-P(k)^{\Lambda C D M}\right) / P(k)^{\Lambda C D M}$. Observations data points have been also shown for comparison.}
    \label{fig:mps}
\end{figure}

The behavior of the background dynamics for the proposed model shows consistent results, but it is important to confirm the model's performance at the perturbation level. To check it, we have computed the temperature anisotropies and the matter power spectrum (MPS) using the CLASS code for the mean value of the parameters obtained in Table \ref {Tab:constraint} for Set 2.  In FIG.\ref{fig:dl} and FIG.\ref{fig:mps} we have shown the plots for the temperature anisotropies and the matter power spectrum (MPS) respectively. To make a comparison with observed data, the values from different observations have been plotted together with the $\Lambda$CDM case for reference. In the FIG.\ref{fig:dl} for temperature anisotropies (TT) the binned TT power spectrum data from Planck 18 ~\cite{Planck:2019nip} has been shown.  For the MPS in FIG.\ref{fig:mps} following are the data sets which have been used: Planck2018 CMB data ~\cite{Planck:2019nip}, SDSS galaxy clustering ~\cite{Reid:2009xm}, SDSS Ly$\alpha$ forest ~\cite{SDSS:2017bih} and DES cosmic shear data ~\cite{DES:2017qwj} (for details on full data collection, see ~\cite{2019MNRAS.489.2247C}).
 
 At the bottom pan of FIG.\ref{fig:dl} and FIG. \ref{fig:mps} we have shown the relative differences in $\Delta D_l=\left(D_l-D_l^{\Lambda C D M}\right) / D_l^{\Lambda C D M}$ and $\Delta P(k)=\left(P(k)-P(k)^{\Lambda C D M}\right) / P(k)^{\Lambda C D M}$ to compare with the $\Lambda CDM$ model. For $D_l$, the proposed model shows a small deviation from $\Lambda CDM$ at higher multipoles, and for $P(k)$ it can be seen at the higher scale. The PPF approximation (Parameterized Post-Friedmann)~\cite{Fang:2008sn} has been used to calculate $D_l$ and $P(k)$. PPF approximation is useful for perturbations to cross the phantom divider smoothly. This approximation has already been implemented in the CLASS code by default. Both the plots in FIG.\ref{fig:dl} and in FIG.\ref{fig:mps} suggest the consistency of the proposed model not only at the background level but also at the perturbation level. 

\section{Comparison with $\Lambda CDM$ and the satus on the Hubble Tension}\label{sec:tension}

\begin{table}
    \centering
    \begin{adjustbox}{width=\columnwidth,center}
    \begin{tabular}{|c|c|c|c|c|c|c|c|}
    \hline
       Model & $\chi^2_{min}$&$\Delta\chi^2$  & $\Delta$AIC & {$\lvert \ln B \rvert $} &{$\lvert \ln B_{w \Lambda} \rvert $} & $T_H$  &  $Q_{DMAP}$   \\
       \hline
        $\Lambda$CDM & 1,342.02 &  0 & 0 & 687.05 & 0 & $ 3.87 \sigma$ & $5.56 \sigma$ \\
        \hline
       $w_q$CDM  & 1,324.52 & -17.5 & -13.5&  682.28 & 5.066 & $2.86 \sigma$  & $ 4.89 \sigma$\\
       \hline
        CPL & 1,327.1 & -14.92 & -10.92& 681.96 & 5.317  & $2.71 \sigma$  & $4.79 \sigma$\\
       \hline
    \end{tabular}
    \end{adjustbox}
    \caption{Best-fit $\chi^2$ values, along with the $\Delta AIC$, $\lvert \ln B_{w \Lambda} \rvert$, Gaussian tension $T_H$, and the $Q_{DMAP}$ for the $\Lambda$CDM, $w_q$CDM, and CPL models for Set 2 combination of the data sets. }
    \label{tab:chisquare}
\end{table}

We use the Bayes factor concept and Akaike Information Criteria (ACI), to determine the preference of this model over the $\Lambda CDM$ model.

Bayes factor is calculated as $\ln B_{w \Lambda} = \ln \mathcal{Z}{w} - \ln \mathcal{Z}\Lambda$, where $\mathcal{Z}$ represents Bayesian evidence and the suffixes $w$ and $\Lambda$ represent the current $wCDM$ and $\Lambda$CDM models, respectively. 

In general, Jeffrey's scale is used to find the preference of one model over another model. According to this scale, a negative preference will be considered if $\lvert \ln B_{w \Lambda}\rvert < 1$, on the other hand positive, moderate, and strong preferences will be considered if $\lvert \ln B_{I \Lambda}\rvert > 1$, $\lvert \ln B_{I \Lambda}\rvert > 2.5$, and $\lvert \ln B_{I \Lambda}\rvert > 5.0$, respectively ~\cite{Trotta:2005ar}. We have used a publicly available code \textsc{MCEvidence}~\cite{Heavens:2017afc} to directly calculate the Bayes factor from the MCMC chains. From Table \ref{tab:chisquare} it can be seen that the evidence for our model is $5.066$, indicating strong evidence for the proposed $w_q$CDM model over the $\Lambda$CDM model, whereas the CPL model also shows similar strong evidence over $\Lambda$CDM.


We also calculated $\Delta\chi^2_{min}$ for the $w_q$CDM and CPL model in comparison with $\Lambda CDM$ which is given in Table \ref{tab:chisquare}. The lowest value of $\Delta\chi^2_{min}$ is obtained for the $w_q$CDM model which is $-17.5$, indicating a better fit to data for the proposed model.

We have also used the concept of the Akaike Information Criterion (AIC) to compare the models against $\Lambda$CDM. The AIC is defined as follows:

\begin{equation}
    \Delta \text{AIC} = \chi^2_{\text{min}, \mathcal{M}} - \chi^2_{\text{min}, \Lambda \text{CDM}} + 2(N_{\mathcal{M}} - N_{\Lambda \text{CDM}})
    \label{eq:placeholder}
\end{equation}

Here the $\mathcal{M}$ and  $N_{\Lambda \text{CDM}}$ represent the number of free parameters of the model $\mathcal{M}$ and the $\Lambda$CDM. In general, a negative $\Delta$AIC when compared to the $\Lambda$CDM model indicates the preference for the model. As it can be seen from Table \ref{tab:chisquare} the lowest $\Delta$AIC corresponds to the $w_q$CDM model suggesting preference for the proposed model over $\Lambda$CDM.

To quantify the status of Hubble tension in the proposed model we have used two different approaches. One is the Gaussian tension estimator and another is the $Q_{DMAP}$ ( difference of the maximum a posteriori) estimator. The estimation of tension we obtained using these two approaches for the $w_q$CDM model is given in Table \ref{tab:chisquare} together with $\Lambda$CDM and CPL models for comparison.
 
 The Gaussian estimator proposed in~\cite{Camarena:2018nbr}, $T_{H 0}=\frac{\left|H_0-H_0^{l o c}\right|}{\sqrt{\sigma_{H 0}^2+\sigma_{\mathrm{loc}}^2}}$, where $T_{H0}$ denotes the estimation of the amount of tension, $H_0$ and $H_0 ^{loc}$ is the mean value of the Hubble parameter obtained from the current analysis and local universe observations, respectively, while $\sigma^2$ represents the corresponding variance of the posteriors. By comparing our results with the result obtained in ~\cite{Riess:2021jrx} $\left(H_0=73.04 \pm 1.04 \mathrm{~km} \mathrm{~s}^{-1} \mathrm{Mpc}^{-1}\right)$, from the Hubble Space Telescope and the SH0ES data, the tension status is $T_{H 0} \simeq$ $2.86 \sigma$. On the other hand, when comparing it with the reported value of $\left(H_0=73.22 \pm\right.$ $2.06 \mathrm{~km} \mathrm{~s}^{-1} \mathrm{Mpc}^{-1}$ ) from the standardized TRGB and Type Ia supernova data sets in ~\cite{scolnic2023cats}, it reduces to $T_{H 0} \simeq 1.69 \sigma$. If we consider the classification of the tension in Table IV in ~\cite{Camarena:2018nbr} for the first case, the tension is moderate and weak for the latter case.
 
Another tensiometer $Q_{DAMP}$ was proposed in \cite{Raveri:2018wln} to check how the inclusion of the data from the $SH0ES$ measurement can impact the fit for a proposed model $\mathcal{M}$ and has been used in \cite{Schoneberg:2021qvd} to test the status of the Hubble tension in for a wide range of models. To estimate it one needs to find the best-fit chi-square between the combined data set with $SH0ES$ observation and without it as $\Delta \chi^2 = \chi^2_{\min, \mathcal{D}+\text{SH0ES}} - \chi^2_{\min, \mathcal{D}}$ and the tension will be estimated as $T=\sqrt{\Delta\chi^2}$. \footnote{One can only estimate the $Q_{DAMP}$ tension for datasets differing by one degree of freedom. Since we have used the Pantheon+SH0ES data for constraining the models to calculate $Q_{DAMP}$ we have used a similar approach to \cite{Teixeira:2024qmw}. We have imposed a Gaussian prior on the calibration of the absolute magnitude $M_B$ of the supernovae Pantheon-plus sample. This is done to exclusively compute $Q_{DAMP}$ for the models, and we have verified that this substitution does not alter the results. The value of the $Q_{DAMP}$  for the $\Lambda$CDM mode we obtained is very similar to the one obtained in the  \cite{Teixeira:2024qmw} (Table III) using the same set of data.} In Table \ref{tab:chisquare} we have listed $Q_{DAMP}$ for $w_q$ CDM model and $\Lambda$ CDM and CPL models. The $Q_{DAMP}$  for the $\Lambda$CDM model is 5.56 $\sigma$ whereas for the $w_q$CDM it is $4.89 \sigma$ showing a slight reduction in the tension. Also, the status of the tension using $Q_{DAMP}$ for the CPL model is very similar to the $w_q$CDM model. From Table. \ref{tab:chisquare} it can be seen that for the $w_q$CDM model, the $\chi_{min}^2$ is lowest for the combination of all data sets indicating that this model fits the data better than others. It can be seen from Fig. \ref{fig:H(z)} and Fig.\ref{fig:eos} that this model mimics the $\Lambda$ CDM at higher redshift and differs from it at lower redshift, improving fitting to the data, hence reducing the tension.

\section{Conclusions} \label{conclusion} 

Recent findings from DESI2024~\cite{DESI:2024mwx} show a preference for the dynamical nature of dark energy, including a potential transition from phantom to quintessence phases in the recent past. In this study, we introduce an alternative parameterization of the dark energy equation of state, motivated by quintom-type scalar field models, yielding results comparable to those reported in DESI2024~\cite{DESI:2024mwx}. One interesting feature of our proposed parameterization is that, while it differs from the CPL model at very low redshift, it reduces to the CPL form at high redshift ($z<1$), whereas at higher redshift this model mimics the cosmological constant.  This means that it could provide a richer understanding of cosmology at lower redshifts compared to that of CPL while maintaining consistency with early-time cosmology. The model uses \textit{trigonometric cosine} and \textit{hyperbolic cosine} functions to represent the equation of state (EoS) for quintessence and phantom nature, respectively.

We used recently released DESI BAO R1 data from \cite{DESI:2024mwx}, Pantheon+ supernova data with and without Chepids
host distance anchor, and a compressed Planck likelihood to constrain the cosmological parameters. Our findings also suggest that the current equation of state (EoS) of dark energy is of the quintessence type, with a possible phantom barrier crossing in the recent past, which aligns with the results of DESI2024\cite{DESI:2024mwx,desi2}. The transition from phantom to quintessence occurs at approximately $z\simeq0.35$. For this model, we also find that the current value of the Hubble parameter is higher than that in $\Lambda$CDM models and Hubble tension is significantly reduced in our model, weakening to approximately $2.8\sigma$ compared to the Hubble Space Telescope and SH0ES data~\cite{Riess:2021jrx}, and to $1.6\sigma$ when compared with standardized TRGB and Type Ia supernova data~\cite{scolnic2023cats}. To evaluate its performance at the linear perturbation level, we plotted the temperature anisotropies and the matter power spectrum (MPS) and compared them to the $\Lambda$CDM model. A slight deviation from the $\Lambda$CDM model was observed at high multipoles and large scales, indicating that our model remains consistent even at the perturbation level. We also calculated the Bayes factor and $\Delta AIC$ to assess the preference of this model over the $\Lambda$CDM model and found a strong preference for our model, indicating the dynamical nature of dark energy over the cosmological constant.

\section{Acknowledgement}
The author acknowledges the use of the Chalawan High Performance Computing cluster, operated and maintained by the National Astronomical Research Institute of Thailand (NARIT).

\appendix

\section{Evolution of the $w_q(z)$ for different choices of the $w_a$ and $w_b$ parameters.}\label{app:wz}

\begin{figure}
    \centering
    \includegraphics[width=\columnwidth]{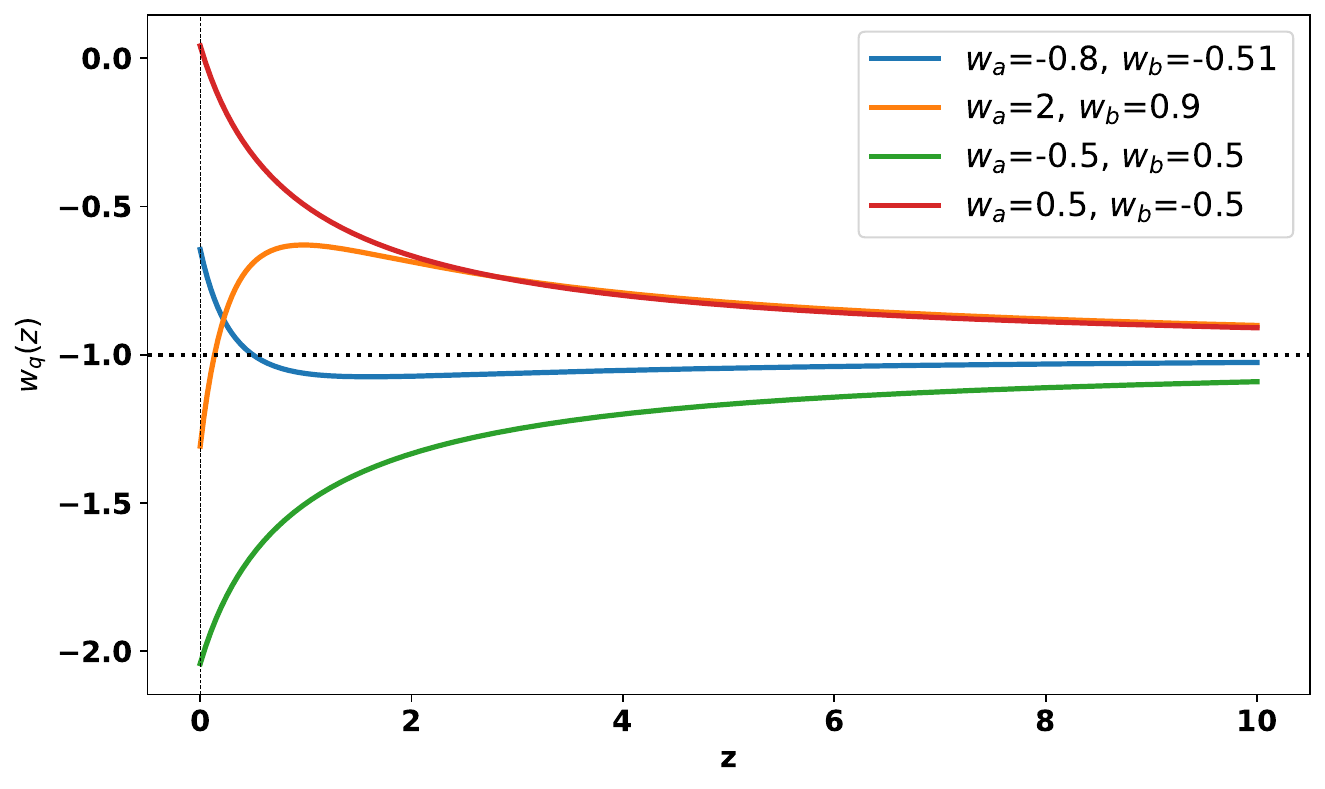}
    \caption{Evolution of $w_q(z)$ with the redshift $z$ for different choices of the $w_a$ and $w_b$ showcasing four different type of scenarios.}
    \label{fig:wqz}
\end{figure}

Here we discuss the evolution of $w_q(z)$ with the redshift $z$ for different choices of the $w_a$ and $w_b$ parameters. As mentioned in Section \ref{sec:mathematicalbg}, this parametrization can mimic the cosmological constant, $w_q \simeq -1$, irrespective of the choice of model parameters as $\frac{a}{a_0} \rightarrow 0$. However, during late times, it may exhibit richer behavior.
In Fig.\ref{fig:wqz}, we have plotted $w_q(z)$ vs $z$ for four different choices of the $w_a, w_b$ parameters, each displaying different qualitative behavior of the $w_q$ evolution at late times. In two cases, $w_q$ evolves from $\simeq -1$ and dips into either the quintessence (in red) or phantom (in green) regions without any phantom barrier crossing. In one case, $w_q$ evolves into the quintessence region and currently transitions to the phantom region through a phantom barrier crossing (in yellow). The plot in blue shows the opposite behavior, where $w_q$ crosses the phantom barrier from the phantom to the quintessence domain. Our choice of priors on the $w_a, w_b$ parameters in this analysis incorporates all these aforementioned cases.

\bibliographystyle{unsrt}
\bibliography{qeos}

\end{document}